!Mode::"TeX:UTF-8"
\UseRawInputEncoding

\documentclass[review]{elsarticle}

\usepackage{lineno}
\usepackage[colorlinks,
            linkcolor=blue,       %%DT???????|?a????????|???????|?
            anchorcolor=blue,  %%DT???????|?a????????|???????|?
            citecolor=blue,        %%DT???????|?a????????|???????|??????????????DT??blue?ared
            ]{hyperref}
\usepackage{amssymb,amsmath}
\journal{}
\usepackage{geometry}
\usepackage{graphicx}
\usepackage{booktabs}
\usepackage{color}
\usepackage{colortbl}
\usepackage{epstopdf}
\usepackage[linesnumbered,ruled,vlined]{algorithm2e}%[ruled,vlined]{
\usepackage{algpseudocode}
\usepackage{amsmath}

\newtheorem{definition}{Definition}
\newtheorem{theorem}{Theorem}
\newtheorem{remark}{Remark}

\newtheorem{example}{Example}
\newtheorem{lemma}{Lemma}
\usepackage{multirow}
\usepackage{graphicx}
\usepackage{subfigure}
\usepackage{cleveref}
\usepackage{titlesec}
\titlespacing*{\section}
{0pt}{0.5ex plus 1ex minus .1ex}{0.1ex plus .1ex}
\titlespacing*{\subsection}
{0pt}{0pt}{0.2pt}
\begin{document}

\begin{frontmatter}

\title{The general conformable fractional grey system model and its applications} %\tnoteref{mytitlenote}}

%\tnotetext[mytitlenote]{Please cite this article as: Jie Xia, Xin Ma, Wenqing Wu, Baolian Huang, Wanpeng Li.
%Application of a new information priority accumulated grey model with time power to predict short-term wind turbine capacity. Journal of Cleaner Production, Volume 244, 2020, 118573, doi: https://doi.org/10.1016/j.jclepro.2019.118573.}%Fully documented templates are available in the elsarticle package on \href{http://www.ctan.org/tex-archive/macros/latex/contrib/elsarticle}{CTAN}.}

%%%%% use optional labels to link authors explicitly to addresses:
\author[label1]{Wanli Xie}
\address[label1]{Institute of EduInfo Science and Engineering, Nanjing Normal University, Nanjing Jiangsu 210097, China}

 \author[label1]{Mingyong Pang\corref{cor1}}
 \cortext[cor1]{Corresponding author. }
 \ead{panion@netease.com}

\author[label3]{Wen-Ze Wu}
 \address[label3]{School of Economics and Business Administration, Central China Normal University, Wuhan 430079, China}

 %%%%%%%%%%%%%%%%%%%%%%%%%%%%%%%%%%%%%%%%%%%%%%%%%%%%%%%%%%%%
\author[label4]{Chong Liu}
 \address[label4]{School of Science, Northeastern University, Shenyang 110819, China}

\author[label1]{Caixia Liu}

%%%%%%%%%%%%%%%%%%%%%%%%%%%%%%%%%%%%%%%%%%%%%%%%%%%%%%%%%%%%%%%%%%%%
%=================================================================
\begin{abstract}
 Grey system theory is an important mathematical tool for describing uncertain information in the real world. It has been used to  solve the uncertainty problems specially caused by lack of information. As a novel theory, the theory can deal with various fields and plays an important role in modeling the small sample problems. But many modeling mechanisms of grey system need to be answered, such as why grey accumulation can be successfully applied to grey prediction model? What is the key role of grey accumulation? Some scholars have already given answers to a certain extent. In this paper, we explain the role from the perspective of complex networks. Further, we propose generalized conformable accumulation and difference, and clarify its physical meaning in the grey model. We use our newly proposed fractional accumulation and difference to our generalized conformable fractional grey model, or GCFGM(1,1), and employ practical cases to verify that GCFGM(1,1) has higher accuracy compared to traditional models.
\end{abstract}
%==========================================================
%=============================================
\end{frontmatter}

%\linenumbers
%%%%%%%%%%%%%%%%%%%%%%%%%%%%%%%%%%%%%%%%%%%%%%%%%%%%%%%
% main text

\section{Introduction}
\label{sec:intro}
Grey system model is a kind of models for modeling uncertain systems and it is also an important mathematical modeling tool to describe the real world \cite{1}. It is very different from the models based on fuzzy mathematics \cite{2} and mathematical statistics \cite{3}. Grey system theory studies the modeling problems under small samples, which allows us to better utilize and process small sample data. Since the establishment of the grey system theory, it has been used to solve many problems in many societies, such as energy \cite{4}, environment \cite{5}, transportation \cite{6}, education \cite{7}, biology \cite{8}, food \cite{9}, chemistry \cite{10}, economics \cite{11}, agricultural science \cite{12}, engineering management \cite{13} and so on, it has now  become an important theory of uncertain systems. As an important mathematical tool, the grey system includes many effective models, such as grey prediction model \cite{14}, grey correlation model \cite{15}, grey decision model \cite{16}, grey programming model \cite{17}, grey game model \cite{18}. Among these models, grey prediction model is a research hotspot, which can solve the problem of poor information successfully. Grey accumulation is an important operator in grey prediction, which has ability to fully expose the hidden information in the raw data \cite{1,19}. As a grey prediction model proposed earlier, GM(1,1) is widely used in various fields \cite{20}. In order to improve the accuracy of the GM(1,1), Xie and Liu proposed the DGM(1,1) model in \cite{61}, which can directly derive the time response sequence by the difference equation. In order to make the model have the ability to fit nonlinear raw data, Chen proposed the grey Bernoulli model in \cite{21}. Recently, Liu and Xie presented a new nonlinear grey model with Weibull cumulative distribution, and gave many valuable conclusions in \cite{22}. Luo and Wei proposed a new grey polynomial model, which has good performance to prediction of time series in \cite{23}. Ma and Liu, Ye and Xie  give two grey polynomial models with time delay effects and achieve good results respectively in \cite{24} and \cite{25}. Recently, the grey Riccati model by Wu has been proposed, which is also a nonlinear grey model and has been successfully applied in energy field \cite{26}. Other pioneering works on grey forecasting models can be found in \cite{28,29,30,31,32}. The models above are very enlightening and are important research results in the grey prediction models. But the order of the models are still fixed to an integer. Some scholars believe that integer-order accumulation is not necessarily optimal. Wu et al. \cite{33} proved that fractional accumulation can reduce the disturbance of the least square solution. Ma et al. \cite{34} proposed a simple and effective method of fractional accumulation and difference, and the method successfully used in the modeling the grey system. Yan et al. \cite{35} considered factional Hausdorff derivative to propose a novel fractional grey model, and gave some valuable results. To further expand the scope of application of the grey model, some researchers considered introducing continuous fractional derivatives into the differential equation of the grey model. Combined with Caputo derivative, Wu et al. \cite{36} earlier proposed a  new grey model. Mao et al.  \cite{37} proposed a new fractional grey prediction model based on the fractional derivative with non-singular exponential kernel. Xie et al. used both conformable fractional difference and conformable fractional derivative in our new fractional grey model in \cite{38}. These fractional grey prediction models have their own characteristics and can be used to solve various problems, however these models rely on fractional calculus, which specially plays an important role in materials \cite{39}, images \cite{40}, medicine \cite{41} and other fields. By extending the classical calculus, many scholars have developed some important fractional calculus formulas, such as Grunwald-Letnikov derivative \cite{42}, Riemann-Liouville \cite{43} derivative, Caputo \cite{44} derivative and so on. In addition, some other new derivatives have also been proposed to solve many practical problems, such as the Caputo and Fabrizio \cite{45} proposed a new fractional derivative with no singular kernel. Atangana and Baleanu \cite{46} further expands it with non-local characteristics. Recently, Khalil et al.\cite{47} proposed a novel derivative called conformable derivative with many properties consistent with classical derivatives. Zhao et al. \cite{48} proposed a class of generalized fractional derivatives and presented the physical explanation. Inspired by \cite{48}, in this paper, we propose a class of generalized fractional difference, accumulation and a new grey prediction model. The rest of the paper is organized as follows: In Section \ref{s2} we explain the advantages of first-order accumulation from the perspective of complex networks, and proposes a generalized conformable accumulation and difference; In Section \ref{s3}, we proposes a onformable fractional grey model and present an optimization method for our model order; In Section \ref{s4}, we shows two concrete cases to verify the effectiveness of the model, and shows the optimization process of the model; In Section \ref{s5}, we draw the conclusion for our method.
\begin{table}[htbp]\scriptsize
  \centering
  \caption{Research results of fractional grey models}
    \begin{tabular}{lllp{21.39em}}
    \toprule
    Author (year) & Abbreviation & Case  & \multicolumn{1}{l}{Description } \\
    \midrule
    \multicolumn{4}{p{45.67em}}{Conformable fractional grey models} \\
    \midrule
    Ma et al. (2019)\cite{34} & CFGM(1,1) & \multicolumn{1}{p{8.055em}}{Simulative case} & The new definitions of conformable fractional accumulation and difference\newline{}are proposed at the first time \\
    Wu et al. (2020)\cite{56} &  FANGM(1,1,k,c) & \multicolumn{1}{p{8.055em}}{Carbon dioxide emissions} & Developing conformable fractional  non-homogeneous grey model with  matrix form of fractional order accumulation operation \\
    Javed et al. (2020)\cite{5} & \multicolumn{1}{p{7.945em}}{EGM(1,1,$\alpha$,$\theta$  )} & \multicolumn{1}{p{10em}}{Biofuel production and consumption} & Designing a novel conformable fractional grey model  with weighted background value  \\
    Xie et al. (2020) \cite{OCFNGM} & CFONGM(1,1,k,c) & Simulative case & Optimizing the background value of the conformable fractional non-homogeneous grey model \\
    Xie et al. (2020)\cite{38} & CCFGM(1,1) & Simulative case & Establishing continuous grey model with\newline{}conformable fractional derivative \\
    Zheng et al. (2021)\cite{MFONGBM} & CFNHGBM(1,1,k) & \multicolumn{1}{p{8.055em}}{Natural gas production and consumption} & Constructing a MFO-based conformable fractional nonhomogeneous grey Bernoulli\newline{}model \\
    Xie et al. (2021) \cite{CCFNGBM} & CCFNGBM(1,1) & \multicolumn{1}{p{8.055em}}{Carbon dioxide emissions} & Optimizing the  nonlinear grey Bernoulli model with conformable fractional derivative \\
    Wu et al. (2021)\cite{WuUFNGBBM} & FDNGBM(1,1) & \multicolumn{1}{p{8.055em}}{ Wind turbine capacity} & Introducing a novel fractional discrete nonlinear grey Bernoulli model   with conformable fractional accumulation \\
    \midrule
    \multicolumn{4}{p{45.67em}}{Other important fractional grey models} \\
    \midrule
    Wu et al. (2013)\cite{33} & FGM(1,1)  & Simulative case  & The concept of  fractional grey forecasting model is put forward at the first time \\
    Yang and Xue (2016)\cite{GL_grey} & GM(q,1)/GM(q,N)  & \multicolumn{1}{p{8.055em}}{Per capita output of electricity} & Establishing continuous fractional grey model based on the observation error feedback \\
    Mao et al. (2016)\cite{62}  & FGM(q,1)  & Simulative case & Constructing a new fractional grey prediction model  with fractional differential equation  \\
    Wu et al. (2019)\cite{FNGBMWU} & FANGBM(1,1) & \multicolumn{1}{p{8.055em}}{Renewable energy consumption} & Establishing a novel fractional nonlinear grey Bernoulli model \\
    Ma et al. (2019)\cite{MAFTDGM} & FTDGM & \multicolumn{1}{p{8.055em}}{Natural gas and coal consumption} & Designing a novel grey model with  fractional time delayed term \\
    \multicolumn{1}{p{10em}}{ Mao et al. (2020)\cite{37}} & FGM(q,1)/PFGM(q,1) & \multicolumn{1}{p{8.055em}}{Electronic waste precious metal\newline{}content} & Establishing a new fractional grey model based on non-singular exponential kernel \\
    \multicolumn{1}{p{10.28em}}{Meng et al. (2020)\cite{MUFGM}} &  FDGM(1,1) & Simulative case & The concept of uniform of fractional grey generation operators is given at the first time \\
    Yan et al. (2020)\cite{35} & FHGM(1,1)  & Simulative case & Establishing a new  fractional  grey model with fractional Hausdorff derivative  \\
    Liu et al. (2021)\cite{Liu}  & DAGM(1,1) & Simulative case & The definition of the damping accumulation is given at the first time \\
    Wu et al. (2021)\cite{Wu_SFGM} & SFNDGM(1,1) & \multicolumn{1}{p{10em}}{Electricity consumption} & Building a novel seasonal fractional nonhomogeneous discrete grey model \\
    Liu et al. (2021)\cite{OFAGM_LIU}  & OFAGM(1,1) & \multicolumn{1}{p{10em}}{Electricity consumption} & Reconstructing a dynamic background value for the fractional grey\newline{}model \\
    Kang et al. (2021)\cite{kang_VFGM} & VOAKFGM  & Simulative case  & Introducing a novel variable order fractional grey model  at the first time \\
    Zeng (2021)\cite{FTDGM_zeng} & NGM(1,1,$\tau$,r) & Energy consumption & Establishing a time delay grey model with fractional order accumulation \\
    \bottomrule
    \end{tabular}%
  \label{tab:addlabel}%
\end{table}%
\section{A class of generalized conformable fractional accumulation}
\label{s2}
% \begin{figure}
% \centering
% \includegraphics[scale=0.35,keepaspectratio]{accuration-1.eps}
% \caption{The predicted values by five competitors in China' domestic coal consumption.}
% \label{fig:accuration_line}
% \end{figure}
In this section, we will first give an analysis of integer-order accumulation based on the theory of complex networks. Secondly, we will propose a new generalized conformable fractional-order accumulation and difference.
\subsection{Understanding of integer order accumulation based on perspective of complex network}
\label{1AGO_network}
\begin{figure}
\centering
\includegraphics[scale=0.4]{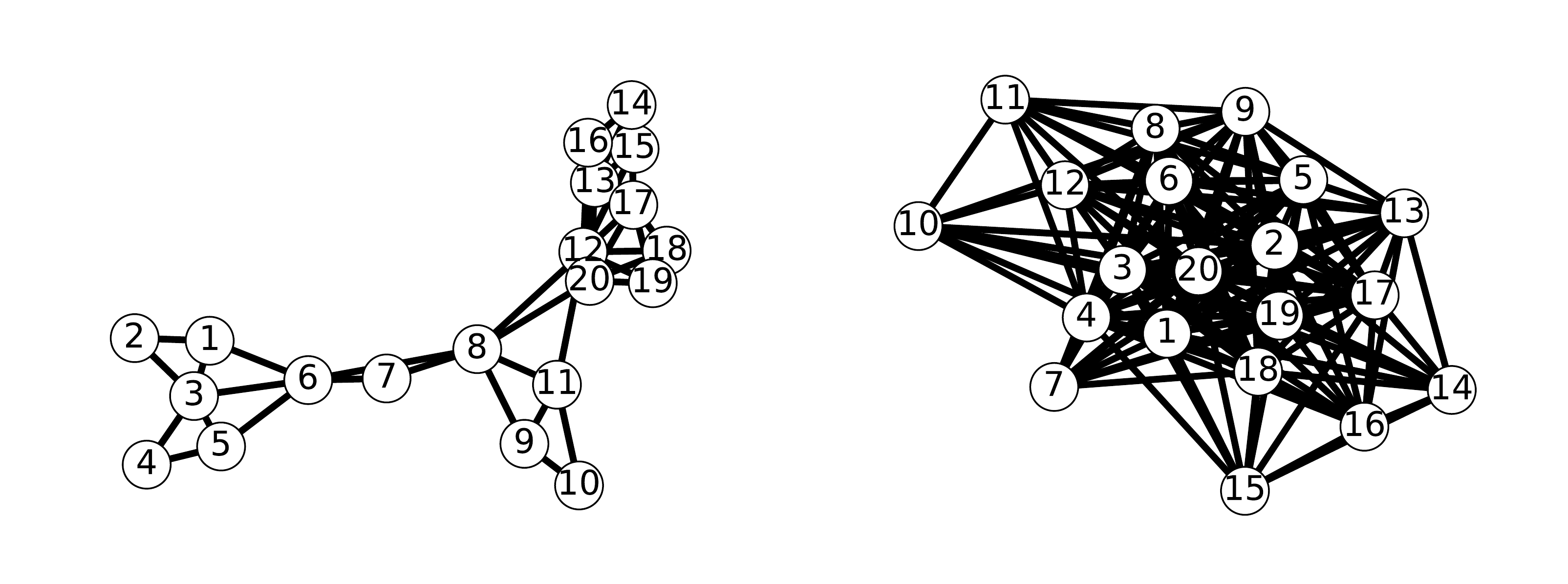}
\caption{Networks corresponding to a original sequence (left) and it first-order accumulation sequence (right).}
\label{fig:graph_one_order}
\end{figure}
There are two types of explanations of grey accumulation and some important research results \cite{19,49}. In this subsection, we explain the advantages of grey accumulation by means of complex network theory. We use the data of inbound tourists (10,000 people) downloaded from the National Bureau of Statistics of China (http://www.stats.gov.cn/) for our explanation. Firstly, we convert the original sequence and the first-order accumulation sequence into the form of a complex network  respectively.
\begin{definition}[See \cite{50}]
\label{graph_to_time}
Suppose $X$ is an original sequence, $X = \left( {\boldsymbol{x}_1^{(0)},\boldsymbol{x}_2^{(0)}, \ldots  \ldots ,\boldsymbol{x}_n^{(0)}} \right)$, and the transformed network is set to ${\rm{G}} = ({\rm{V}},{\rm{E}})$, if $\forall \boldsymbol{x}_a^{(0)}$ and $\boldsymbol{x}_b^{(0)}$, $\exists {\rm{\boldsymbol{x}}}_c^{(0)}$, makes
\begin{equation} \label{eqU:ttf}
\boldsymbol{x}_c^{(0)} < \boldsymbol{x}_b^{(0)} + \left( {\boldsymbol{x}_a^{(0)} - \boldsymbol{x}_b^{(0)}} \right)\frac{{{t_b} - {t_c}}}{{{t_b} - {t_a}}};({t_a} < {t_c} < {t_b}),
\end{equation}
where $\left( {{t_c},x_c^{(0)}} \right)$ is a point between $ \left( {{t_a},x_a^{(0)}} \right) and \left( {{t_b},x_b^{(0)}} \right)$, then there exists an edge between $x_a^{(0)}$ and $x_b^{(0)}$. Through \textcolor {blue}{Eq.} (\ref{eqU:ttf}), we can find that if $\left( {{t_c},x_c^{(0)}} \right)$ is the largest number of $\left( {{t_a},x_a^{(0)}} \right)$ and $\left( {{t_b},x_b^{(0)}} \right)$, then a and b cannot have a link relationship, that is, when there are more fluctuations in the original time series, there will be fewer links in the corresponding network. However, when the first order accumulating generation operator (1?AGO) is employed to preprocess the original data, the sequence is strictly monotonically increasing. As long as $\left( {{t_b},x_b^{(0)}} \right)$ is large enough, there may be a link between $\left( {{t_a},x_a^{(0)}} \right)$ and $\left( {{t_b},x_b^{(0)}} \right)$. This means that the network formed by the 1-AGO series has more chances to have connections than the original series.
\end{definition}
According to \textcolor{blue}{Definition} \ref{graph_to_time}, we map the time series into a complex network.
In Figure \ref{fig:graph_one_order}, we show the complex network after the conversion of the original data and 1-AGO series. We then respectively calculated two types of statistical indicators of the network, namely the clustering coefficient \cite{51} and the average path length \cite{52}. The definition of first-order grey accumulation \cite{1}, clustering coefficient, and average path length are as
\begin{equation}
% \label{gmp_b1}
{x^{(1)}}(k) = \sum\limits_{s = 1}^k {{x^{(0)}}} (s),...n, C = \frac{1}{n}\sum\limits_{i = 1}^n {\frac{{2{E_i}}}{{{k_i}({k_i} - 1)}}}, \frac{1}{{n(n - 1)}}{\Sigma _{i \ne j}}{d_{ij}},
\end{equation}
\begin{table}\scriptsize
  \centering
  \caption{The original sequence and the first-order cumulative sequence form a network corresponding to
clustering coefficient and average path length.}
    \begin{tabular}{lrr}
    \toprule
    Data  & \multicolumn{1}{l}{Clustering coefficient} & \multicolumn{1}{l}{Average path length} \\
    \midrule
    Original sequence & 0.708016 & 2.668421 \\
    \midrule
    First-order accumulation & 0.856374 & 1.215789 \\
    \bottomrule
    \end{tabular}%
  \label{tab:addlabel}%
\end{table}%
where $\frac{{2{E_i}}}{{{k_i}\left( {{k_i} - 1} \right)}}$ is the ratio of the number of edges ${{E_i}}$ between the node ${{k_i}}$ to the total number of edges. ${d_{ij}}$ refers to the number of edges on the shortest path connecting two nodes, $i$ and $j$. It can be seen that the 1-AGO series have a larger clustering coefficient and a smaller average path length than the original sequence, so relatively speaking, the 1-AGO series have the characteristics of small-world network \cite{53}. This is mainly because the 1-AGO sequence is more closely connected. In the small-world network, the ability of information dissemination and computing, etc. have been enhanced, that is, the network structure corresponding to the accumulation of the original sequence is compact, which has a stronger efficiency of information dissemination.
\subsection{General conformable fractional accumulation and difference}
\begin{definition}[See \cite{48}]
Set that $D_\psi ^pf(u)$ denotes the general conformable derivative of function $f$, which is defined as
\begin{eqnarray}
\label{eq:cd}
D_\psi ^pf(u) = D_\psi ^{p - n}{D^n}f(u) = \mathop {\lim }\limits_{\epsilon \to 0} \frac{{{f^{(n)}}(u + \epsilon \psi (u,p - n)) - {f^{(n)}}(u)}}{\epsilon},
\end{eqnarray}
where $u > 0$, $p  \in (n,n+1]$. Additionally, if $\alpha  \in (0, 1]$, Eq. (\ref{eq:cd}) can be changed to
\begin{equation}
D_\psi ^pf(u) = \mathop {\lim }\limits_{\epsilon \to 0} \frac{{f(u + \epsilon \psi (u,p)) - f(u)}}{\epsilon}
\end{equation}
where  ${\psi (u,p)}$ is a fractional conformable function \cite{48}.
\label{def:4}
\end{definition}
\begin{remark}
When $\psi(u, p)=1$, $D_\psi ^pf(u)$ degenerates to the first order derivative case.
\end{remark}
\begin{remark}
When $\psi (u,p) = {t^{\left\lceil p \right\rceil  - p}}$ and $\alpha \in(n, n+1]$, $D_{\psi}^{p} f(u)$ is equivalent to the Khalil's fractional derivative with arbitrary order \cite{47}, because of $\psi (u,p) = {u^{\left\lceil p \right\rceil  - p}} = \psi (u,p - n) = {u^{\left\lceil {p - n} \right\rceil  - (p - n)}}$. Specially, when $\psi(u, p)=u^{1-p}$ and $p \in \left( {0,1} \right], D_{\psi}^{p} f(u)$ coincides with the Khalil's fractional derivative \cite{47}.
\end{remark}
\begin{remark}
\label{r3}
When $q \in(0,1]$, $\psi(u, p)$ satisfies $\psi(u, 1)=1, \psi(\cdot, p) \neq \psi(\cdot, q)$, where $ p \neq q$. For example, take linear function: $\psi (u,p) = p$ and power function: $\psi(u, p)=p^{2}$, exponent function: $\psi (u,p) = {a^{(1 - p)\xi (p)}}$ \cite{48}.
\end{remark}

\begin{figure}
\centering
\includegraphics[scale=0.3]{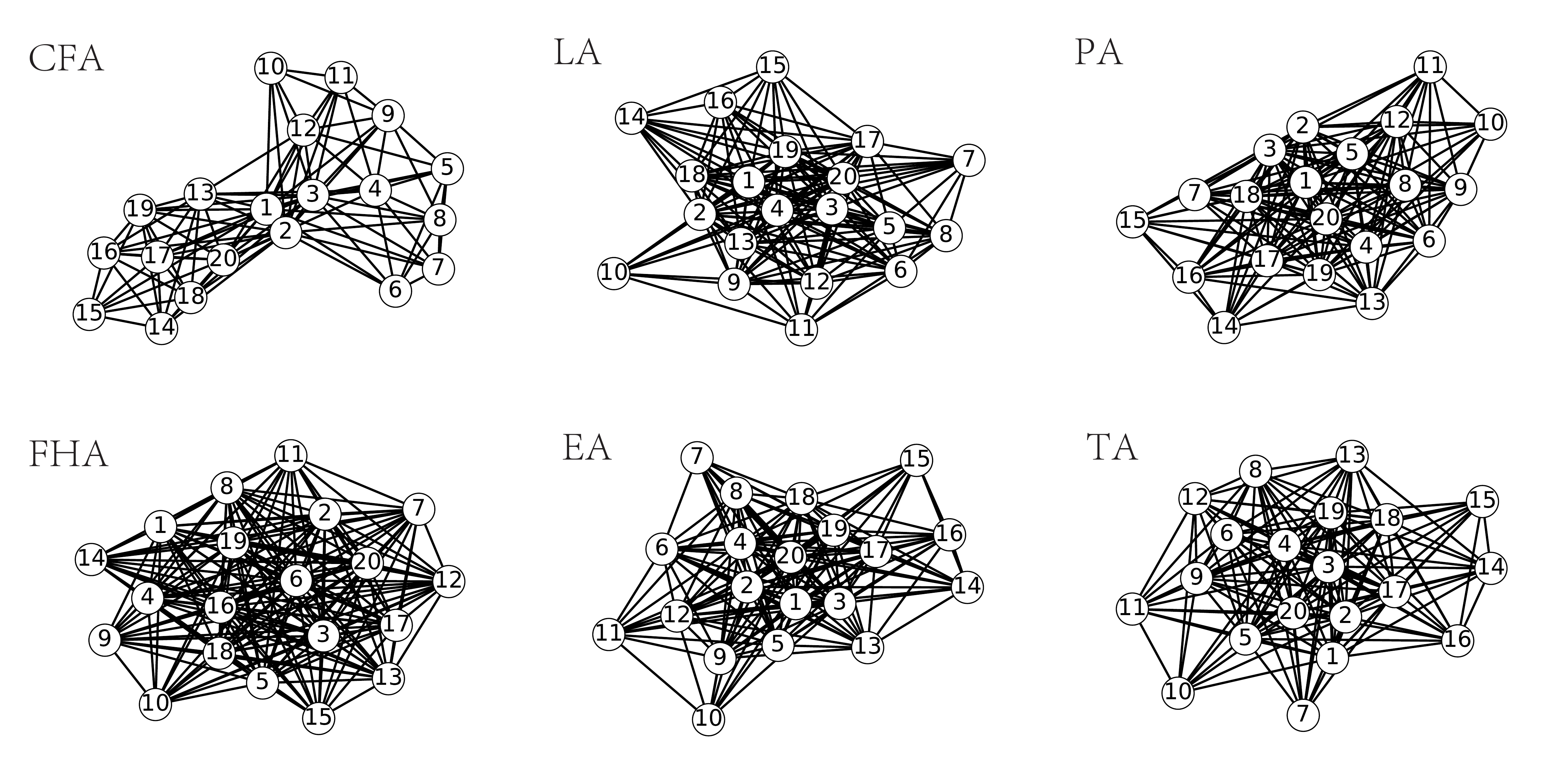}
\caption{Complex networks mapped by different fractional order accumulation.}
\label{fig:graph_FA_order}
\end{figure}

\begin{theorem}[See \cite{48}]
\label{theo:1}
If $f$ is differentiable and ${\rm{ }}t > 0, p \in (0,1]$. Then
\begin{equation}
\label{pro1}
D_\psi ^p(f) = \frac{{df(u)}}{{du}}\psi (u,p)
\end{equation}
{\textbf{ Proof.}} Set $\xi  = \epsilon \psi (u,p)$, then $\epsilon= \frac{\xi }{{\psi (u,p)}}$, therefore
\begin{equation}
\label{transform_order}
\begin{array}{l}
D_\psi ^p(f) = \mathop {\lim }\limits_{\epsilon \to 0} \frac{{f\left( {t + \epsilon \psi (u,p)} \right) - f(t)}}{\epsilon}\\
{\rm{ = }}\psi (u,p)\mathop {\lim }\limits_{\xi  \to 0} \frac{{f(u + \xi ) - f(u)}}{\xi }\\
 = \psi (u,p)\frac{{df(u)}}{{du}}
\end{array}
\end{equation}
\end{theorem}
\begin{remark}
When $p \in \left( {n,n + 1} \right]$, $\psi (u,p)\frac{{{d^{n + 1}}f(u)}}{{d{u^{n{\rm{ + 1}}}}}}$ is also established in \cite{48}.
\end{remark}
According to \textcolor{blue}{Definition} \ref{def:4}, \textcolor{blue}{Theorem} \ref{theo:1}, and the definition of first-order difference ${\left. {\Delta f(k) \approx \mathop {\lim }\limits_{\xi  \to 1} \frac{{f(t) - f(t - \xi )}}{\xi}} \right|_{t = k}} = f(k) - f(k - 1)$ \cite{34}. We discretize the first derivative in \textcolor{blue}{Eq. (\ref{pro1})} into the first difference form, we give the definition of general conformable fractional difference.
\begin{definition} The general conformable fractional difference (GCFD) of $f$ with $\alpha$ order is
\begin{equation}
\begin{array}{*{20}{l}}
{\Delta ^\alpha}f(k) = \psi (k,\alpha)\Delta f(k) = \psi (k,\alpha)[f(k) - f(k - 1)],\alpha \in (0,1],k \in {N^ + }
\end{array}
\end{equation}
\end{definition}
\begin{remark}
When $\psi(k, \alpha)=1$, ${\Delta ^\alpha}f(k)$ degenerates to the first order difference.
\end{remark}
\begin{remark}
When $\psi(k, \alpha)=k^{1-\alpha}, {\Delta ^\alpha}f(k)$ coincides with the Ma's definition of difference \cite{34}.
\end{remark}
\begin{remark}
When $\psi(k, \alpha)=\frac{1}{{{k^\alpha } - {{(k - 1)}^\alpha }}}, {\Delta ^\alpha}f(k)$ coincides with the Yan's definition of difference \cite{35}.
\end{remark}
\begin{example}
Set ${\text{F(k) = (f(1),f(2),f(3)}},{\text{f(4)}},{\text{f(5),f(6)) = (3,7,8}}{\text{.5,12,20,32)}}$, then
\begin{equation}
{\Delta ^\alpha }F(k) = {\text{(}}{\Delta ^\alpha }{\text{f(k))}} = (3,4\psi (2,\alpha ),1.5\psi (3,\alpha ),3.5\psi (4,\alpha ),8\psi (5,\alpha ),12\psi (6,\alpha )),k = 1,2,...,n.
\end{equation}
\end{example}
As described in \cite{34}, integral order difference and accumulation are inverse operations of each other, as shown below,
\begin{equation}
\begin{array}{l}
\Delta \nabla f(k) = \Delta \left( {\sum\limits_{i = 1}^k f (i)} \right) = \sum\limits_{i = 1}^k f (i) - \sum\limits_{i = 1}^{k - 1} f (i) = f(k)
\end{array}
\end{equation}
Inspired by this idea, the GCFD and conformable fractional accumulation (GCFA) should also be inverse operation of each other. It is not difficult to prove that the GCFA and GCFD are inverse operations of each other.
\begin{definition}
The general conformable fractional accumulative (GCFA) sequence with order $\alpha$ is given by
\begin{equation}
\begin{array}{l}
{\nabla ^\alpha }f(k) = \sum\limits_{i = 1}^k {\left( {\begin{array}{*{20}{c}}
{k - i + \lceil \alpha  \rceil - 1}\\
{k - i}
\end{array}} \right)} \frac{{f(i)}}{{\psi (i,\alpha )}},
\label{eq:cfa}
\end{array}
\alpha  \in {R^ + },k \in {Z^ + },
\end{equation}
where $\lceil \alpha  \rceil$  is the smallest integer greater than or equal to $\alpha$,
 $
\binom{k-i+\lceil \alpha  \rceil-1}{k-i}=\frac{(k-i+\lceil \alpha  \rceil-1)!}{(k-i)!(\lceil \alpha  \rceil-1)}$.
\end{definition}
\begin{remark}
When $\psi (i,\alpha ) = 1$, GCFA degenerates into a first-order accumulation.
\end{remark}
\begin{remark}
When $\psi (i,\alpha ) = \alpha^{i-1}$, GCFA coincides with Liu's definition \cite{Liu}.
\end{remark}
\begin{remark}
When $\psi (i,\alpha ) = {i^{1 - \alpha }}, \alpha  \in \left( {0,1} \right],k \in {Z^ + }$,and $\psi (i,\alpha ) = {i^{\left\lceil \alpha  \right\rceil  - \alpha }}, \alpha  \in \left( {n,n+1} \right],k \in {Z^ + }$,  GCFA coincides with Ma's definition \cite{34}. There is also a unified form of CFA \cite{56}.
\end{remark}
\begin{remark}
When $\psi (i,\alpha ) = {i^\alpha } - {(i - 1)^\alpha }$, GCFA is equivalent to the definition of  Yan (In the text below, we call it FHA) \cite{35}.
\end{remark}
\begin{remark}
When $\psi (i,\alpha )$ is a linear function or a power function or an exponential function or a trigonometric function, we call it LA, PA, EA or TA. In short, as long as a meets the requirements in \textcolor{blue}{Remark} \ref{r3}, it is valid.
\end{remark}
We calculate the clustering coefficients and the  average length of different accumulations. We choose $\psi (i,\alpha )$ = $\alpha$ for LA, $\psi (i,\alpha )$ = ${\alpha ^2}$ for PA, $\psi (i,\alpha )$ = ${2^{1 - \alpha }}$ for EA and $\psi (i,\alpha )$, $\sin \left( {\frac{\pi }{2}\alpha } \right)$ for TA with order of 0.9. The statistical indicators of several types of accumulation are listed in the \textcolor{blue}{Table} \ref{FA_network_para}, data from \textcolor{blue}{Subsection} \ref{1AGO_network}. We can see that in this example, the clustering system of FHA is the largest one with a value of 0.933576, and the clustering coefficient of CFA is the smallest, which is 0.782559. They are both greater than the clustering coefficient of the original sequence: 0.708016. The average path length of FHA is the smallest one with 1.08421, and the average path length of CFA is the largest with 1.473684. The network structure formed by different accumulations is illustrated in \textcolor{blue}{Figure} \ref{fig:graph_FA_order}. It can be seen that their structure are different. Further, compared to the original sequence, their structure is more compact.
\begin{table}\scriptsize
  \centering
  \caption{Clustering coefficient and average path length of different fractional accumulation forms with order is 0.9.}
    \begin{tabular}{lrrrrrr}
    \toprule
    Data  & \multicolumn{1}{l}{ CFA} & \multicolumn{1}{l}{ LA } & \multicolumn{1}{l}{PA} & \multicolumn{1}{l}{EA} & \multicolumn{1}{l}{TA } & \multicolumn{1}{l}{FHA} \\
    \midrule
    Clustering coefficient & 0.782559 & 0.856374 & 0.856374 & 0.856374 & 0.856374 & 0.933576 \\
    \midrule
    Average path length & 1.473684 & 1.215789 & 1.215789 & 1.215789 & 1.215789 & 1.08421 \\
    \bottomrule
    \end{tabular}%
  \label{FA_network_para}%
\end{table}%
\section{Generalized conformable fractional grey model}
\label{s3}
In this section, we propose a new grey prediction model based on GCFA and GCFD operators.
\subsection{Basic definition of generalized conformable fractional grey model}
\begin{definition}
With the data sequence ${\bf{{X}}}_{n \times 1}^{(0)} = {\left( {{{x}^{(0)}}(1),{{x}^{(0)}}(2), \ldots ,{{x}^{(0)}}(n)} \right)^T}$, GCFA can be given by ${\bf{{X}}}_{{\rm{n}} \times 1}^{(\alpha)} = {\left( {{{x}^{(\alpha)}}(1),{{x}^{(\alpha)}}(2), \ldots ,{{x}^{(\alpha)}}(n)} \right)^T}$, where
\begin{equation}
{{x}^{(\alpha )}}(k) = {\nabla ^\alpha }{{x}^{(0)}}(k) = \sum\limits_{i = 1}^k {\left( {\begin{array}{*{20}{c}}
{k - i +  \lceil \alpha  \rceil  - 1}\\
{k - i}
\end{array}} \right)} \frac{{{x}(i)}}{{ \psi (i,\alpha )}}, k=1,2,3,...,n.
\label{GCFA}
\end{equation}
We represent the $p$-order ($p \in \left( {0,1} \right]$)differential equation of general conformable fractional grey model GCFGM(1,1) wirth the $\alpha$-GCFA (\textcolor{blue}{Eq.} (\ref{eq:cfa})) series as
\begin{eqnarray}
D_\psi ^p{{x}^{(\alpha )}}(t) + {a}{{x}^{(\alpha )}}(t) = {b}.
\label{eq:ccfde}
\end{eqnarray}
Obviously, when $p=1$ and $\alpha=1$, the model degenerates to GM(1,1) \cite{1}. In the actual modeling environment, the appropriate accumulation method should be selected according to the actual background of data. In particular, the accumulation can be also choosed the weighted form of two functions, such as
\begin{eqnarray}
\label{suggestion_accumulation}
{\nabla ^\alpha }{{x}^{(0)}}(k) = \sum\limits_{i = 1}^k {\left( {\begin{array}{*{20}{c}}
{k - i +  \lceil \alpha  \rceil  - 1}\\
{k - i}
\end{array}} \right)} \frac{{{x}(i)}}{{\frac{1}{2}\left( {{\alpha ^{i - 1}}{\rm{ + }}{i^{1 - \alpha }}} \right)}}, k=1,2,3,...,n.
\label{Recomand_GCFA}
\end{eqnarray}
\end{definition}
\begin{theorem} \label{theorem:exact_solution}
The exact solution to Eq. (\ref{eq:ccfde}) is
\begin{eqnarray}
\label{GCCF_time_response}
{x^{(\alpha )}}(t) = {{\rm{e}}^{ - \int {\frac{a}{{\psi (t,p)}}} {\rm{dx}}}}\left( {\int {\frac{b}{{\psi (t,p)}}} {{\rm{e}}^{\smallint \frac{a}{{\psi (t,p)}}{\rm{dx}}}}{\rm{dx}} + {\rm{C}}} \right).
\end{eqnarray}
\end{theorem}
\textbf{Proof.} Using \textcolor{blue}{Eq.} (\ref{transform_order}), \textcolor{blue}{Eq.} (\ref{eq:ccfde}) can be arranged as
\begin{eqnarray} \label{to_one_order}
\psi (t,p)\frac{{{\rm{d}}{x^{(\alpha )}}(t)}}{{{\rm{d}}t}} + a{x^{(\alpha )}}(t) = b.
\label{eq:p1}
\end{eqnarray}
Set ${x^{(\alpha )}}(x) = C(x) \cdot {e^{ - \int {\frac{a}{{\psi (t,p)}}} dx}}$, we have
\begin{eqnarray} \label{trans_equ}
\frac{{d{x^{(\alpha )}}(x)}}{{dx}} = \frac{{dC(x)}}{{dx}}{e^{ - \int {\frac{a}{{\psi (t,p)}}} dx}} - {x^{(\alpha )}}(x)\frac{a}{{\psi (t,p)}}.
\end{eqnarray}
Based on \textcolor {blue}{Eq.} (\ref{to_one_order}), we have
\begin{eqnarray}  \label{raw_equ}
\frac{{d{x^{(\alpha )}}(x)}}{{dx}} = \frac{b}{{\psi (t,p)}} - \frac{a}{{\psi (t,p)}}{x^{(\alpha )}}(x).
\end{eqnarray}
Combining \textcolor {blue}{Eq.} (\ref{eq:ccfde}) and \textcolor {blue}{Eq.} (\ref{raw_equ}), we have
\begin{eqnarray}  \label{equ:trans}
\frac{{dC(x)}}{{dx}} = {e^{\int {\frac{a}{{\psi (t,p)}}} dx}}\frac{b}{{\psi (t,p)}}, C\left( x \right) = \int {\frac{b}{{\psi (t,p)}}} {e^{\smallint \frac{a}{{\psi (t,p)}}dx}}dx + {C_1}.
\end{eqnarray}
 Substitite \textcolor {blue}{Eq.} (\ref{equ:trans}) into ${x^{(\alpha )}}(x) = C(x) \cdot {e^{ - \int {\frac{a}{{\psi (t,p)}}dx} }}$, we can  get \textcolor {blue}{Eq.} (\ref{GCCF_time_response}). Set ${x^{(\alpha )}}(1) = {x^{(0)}}(1)$, we can get the time response sequence of GCFGM(1,1)by \textcolor{blue} {Theorem} \ref{theorem:exact_solution} as
\begin{eqnarray}
{x^{(\alpha )}}(t) = {{\rm{e}}^{ - \int {\frac{a}{{\psi (t,p)}}} {\rm{d}}x}}\left( {\int {\frac{b}{{\psi (t,p)}}} {{\rm{e}}^{\int {\frac{a}{{\psi (t,p)}}{\rm{dx}}} }}{\rm{dx}} + \frac{{{x^{(0)}}(1)}}{{{{\rm{e}}^{ - \int {\frac{a}{{\psi (t,p)}}} {\rm{d}}x}}}} - \int {\frac{b}{{\psi (t,p)}}} {{\rm{e}}^{\int {\frac{a}{{\psi (t,p)}}{\rm{dx}}} }}{\rm{dx}}} \right).
\label{trespose1}
\end{eqnarray}
In order to estimate the parameters $[\hat{a}, \hat{b}]^{T}$ in GCFM(1,1), we need to discretize \textcolor{blue}{Eq.} (\ref{eq:ccfde}).
Integrating both sides of Eq. (\ref{eq:ccfde}) with $p$ order, we have
\begin{eqnarray}
\iint  \cdots \int_{k - 1}^k {\frac{{{d^p}{x^{(\alpha)}}}}{{d{t^p}}}} d{t^p} + a\iint  \cdots \int_{k - 1}^k {{x^{(\alpha)}}} (t)d{t^p} = b\iint  \cdots \int_{k - 1}^k d {t^p}.
\label{eq:lr}
\end{eqnarray}
Taking the p-th order integral on $\frac{{{d^p}{x^{(\alpha )}}(t)}}{{d{t^p}}}d{t^p}$, we can get
\begin{eqnarray}
\iint  \cdots \int_{k - 1}^k {\frac{{{d^p}{x^{(\alpha)}}(t)}}{{d{t^p}}}} d{t^p} \approx \Delta ^px^{(\alpha)}(k) = {x^{(\alpha - p)}}(k).
\label{eq:l1}
\end{eqnarray}
Using the general trapezoid formula \cite{62}, we can get
\begin{eqnarray}
a\iint  \cdots \int_{k - 1}^k {{x^{(\alpha)}}} (t)d{t^r} \approx \frac{a}{{\rm{2}}}\left( {{x^{(\alpha)}}(k - 1) + {x^{(\alpha)}}(k)} \right).
\label{eq:l2}
\end{eqnarray}
The right-hand side of Eq. (\ref{eq:ccfde}) can be obtained by Ref. \cite{62}, as
\begin{eqnarray}
\iint  \cdots \int_{k - 1}^k b d {t^r}=b \iint  \cdots \int_{k - 1}^k d {t^r}\approx \int_{k - 1}^k {bd} t \approx b.
\label{eq:r1}
\end{eqnarray}
Substituting Eqs. (\ref{eq:l1})-(\ref{eq:r1}) into Eq. (\ref{eq:lr}), We can get the discrete form of the GCFGM model as follows,
\begin{eqnarray}
{{x}^{(\alpha  - p)}}(k) + \frac{{a}}{2}\left( {{{x}^{(\alpha )}}(k) + {{x}^{(\alpha )}}(k - 1)} \right) = {{b}}, k=1,2,3,...,n.
\end{eqnarray}
As we know, the differential equations and corresponding difference equations of the classic GM(1,1) model are as follows,
\begin{eqnarray}
\frac{{d{{x}^{(1)}}(t)}}{{dt}} + {a}{{x}^{(1)}}(t) = {b},{{x}^{(0)}}(k) + \frac{{a}}{2}\left( {{{x}^{(1)}}(k) + {{x}^{(1)}}(k - 1)} \right) = {b}, k=1,2,3,...,n.
\end{eqnarray}
It is well known that $\frac{{d{x^{(1)}}(t)}}{{dt}}$ is an continuous representation of ${x^{(0)}}(k) = {x^{(1)}}(k) - {x^{(1)}}(k - 1)$. Therefore, the effect of the first-order derivative can be approximately regarded as a bridge from the first-order cumulative generated sequence to the original sequence. The physical meaning of the classical first derivative is very clear, which means velocity of particle or slope of a tangent respectively \cite{48}. But due to the uncertainty and complexity of the real world, in the grey system, the change from the first-order cumulative sequence to the original sequence does not necessarily satisfy the law of classical derivatives. Zhao and Luo \cite{48} gave the physical meaning of the general conformable fractional derivative: GCFD is a modification of classical derivative in direction and magnitude. Therefore, in the grey system model, we use GCFD for modeling real world system, which represents a special change from the first-order cumulative sequence to the original sequence. The parameters of the GCFGM model can be obtained by the least squares method. Set ${z^{(\alpha )}}(k + 1) = \frac{{{x^{(\alpha )}}(k) + {x^{(\alpha )}}(k + 1)}}{2},k = 1,2, \cdots ,n - 1$, we have
\begin{eqnarray}
[\hat a,\hat b] = \mathop {\arg \min }\limits_{a,b} \left\{ {\sum\limits_{i = 1}^{n - 1} {{{\left[ {{x^{(\alpha  - p)}}(k) - \left( { - a \cdot {z^{(\alpha )}}(k + 1) + b} \right)} \right]}^2}} } \right\} = {\left( {{B^T}B} \right)^{ - 1}}{B^T}Y,
\label{para1}
\end{eqnarray}
where
$$
B = \left( {\begin{array}{*{20}{c}}
  { - {z^{(\alpha )}}(2)}&1 \\
  { - {z^{(\alpha )}}(3)}&1 \\
   \vdots & \vdots  \\
  { - {z^{(\alpha )}}(n)}&1
\end{array}} \right),Y = \left( {\begin{array}{*{20}{c}}
  {{x^{(\alpha - p)}}(2)} \\
  {{x^{(\alpha - p)}}(3)} \\
   \vdots  \\
  {{x^{(\alpha - p)}}(n)}
\end{array}} \right).
$$
Using the definition of GCFD, the restored values can be written
\begin{eqnarray}
{{{\hat x}^{(0)}}(k) = \psi (k,\alpha )\left( {{{\hat x}^{(\alpha )}}(k) - {{\hat x}^{(\alpha )}}(k - 1)} \right).k = 2,3, \ldots ,n.}
\label{GCFD}
\end{eqnarray}
Through the model established above, we can get the result of GCFGM $\widehat{X}^{(0)}$. In the above analysis, we assume that the order of the model $p$ and $\alpha$ are already known. But in practice, the parameters should be dynamically adjusted according to the actual data. In order to better understand the modeling process of GCFGM, the modeling steps of GCFGM with given time series and $\alpha$ and p can be summarized as follows:

$\pmb{Step\ 1:}$  Calculate the $\alpha$ order GCFA sequences $\left( {{x^{(\alpha )}}(1),{x^{(\alpha )}}(2), \ldots ,} \right.\left. {{x^{(\alpha )}}(n)} \right)$ of the raw series

 $\left(\chi^{(0)}(1), x^{(0)}(2), \ldots, x^{(0)}(n)\right)$;

$\pmb{Step\ 2:}$ Calculate the parameters of GCFGM $\hat{a}$ and $\hat{b}$ by \textcolor{blue}{Eq. (\ref{para1})}

$\pmb{Step\ 3:}$ Calculate the predicted values of the GCFA series using \textcolor{blue}{Eq. (\ref{trespose1})};

$\pmb{Step\ 4:}$ Calculate the restored values of the GCFD series $\hat{x}^{(0)}(k),k=1,2,3,..,n+u$  using \textcolor{blue}{Eq. (\ref{trespose1})}, where n is the number of
samples for building model and u is the number of prediction steps;

\subsection{Intelligent optimization algorithm for selecting the optimal p and $\alpha$}
In the above analysis, we assume that the order of GCFM is known. So, in order to get the appropriate order, we can design the following model,
\begin{eqnarray}
{\min _{\alpha,p}, }{\text{MAPE }} = \sum\limits_{i = 1}^{\text{n}} {\left| {\frac{{{{\hat x}^{(0)}}(i) - {x^{(0)}}(i)}}{{{x^{(0)}}(i)}}} \right|}  \times 100\% ,{\text{ s}}{\text{.t}}{\text{. }}\left\{ \begin{gathered}
  \textcolor{blue}{Eq}.(\ref{GCFA}) \hfill \\
  \textcolor{blue}{Eq}.(\ref{para1}) \hfill \\
  \textcolor{blue}{Eq}.(\ref{trespose1}) \hfill \\
   \textcolor{blue}{Eq}.(\ref{GCFD}) \hfill \\
\end{gathered}  \right. .
\label{fitness1}
\end{eqnarray}
In order to get an appropriate order, combined with the above model, We first consider using PSO\cite{60} search for the order of GCFM model respectively, which are widely used in engineering and science, and have achieved good performance. The concrete algorithm for searching order of model is presented in Algorithm \textcolor{blue}{1}.
\begin{algorithm}\scriptsize
  \caption{Particle swarm optimization to search the fractional
order $\alpha$ and $p$ of GCFGM}
  \KwIn{Raw data $X ^{( 0 )}=\left\{x^{(0)}(1), x^{(0)}(2), \ldots, x^{(0)}(n)\right\}$}
  \KwOut{Optimized fractional order $\ \alpha, p$}
  \textbf{Set} the maximum iteration number and maximum population

  \textbf{Initialize} Particle swarm

   \textbf{Initialize} velocity and position

    \For{$t \leftarrow 1;t < {\text{max iteration number}};{\text{t}} \leftarrow {\text{t + 1}}$}{
    \For{$j \leftarrow 1;j < {\text{maximum population}};{\text{j}} \leftarrow {\text{j + 1}}$}{
    Update
    the velocity and position

       Construct $\alpha$-GCFA series of $X^{(0)}$ using \textcolor{blue}{Eq}.(\ref{GCFA})

        Get the parameters of the model $[\hat{a}, \hat{b}]$ using   \textcolor{blue}{Eq}.(\ref{para1})

        Compute $\hat{X}^{(r)}(k)$ by time response function of GCFGM using  \textcolor{blue}{Eq}.(\ref{trespose1})

        Compute restored values $\hat{x}^{(0)}(k)$ using \textcolor{blue}{Eq}.(\ref{GCFD})

        Compute fitness using  \textcolor{blue}{Eq}.(\ref{fitness1})
      }
     }
     \textbf{Return} the optimized order $\alpha$
  \label{algorithmone}
\end{algorithm}
\subsection{Properties of the GCFGM model}
\begin{figure}
\centering
\includegraphics[scale=0.72]{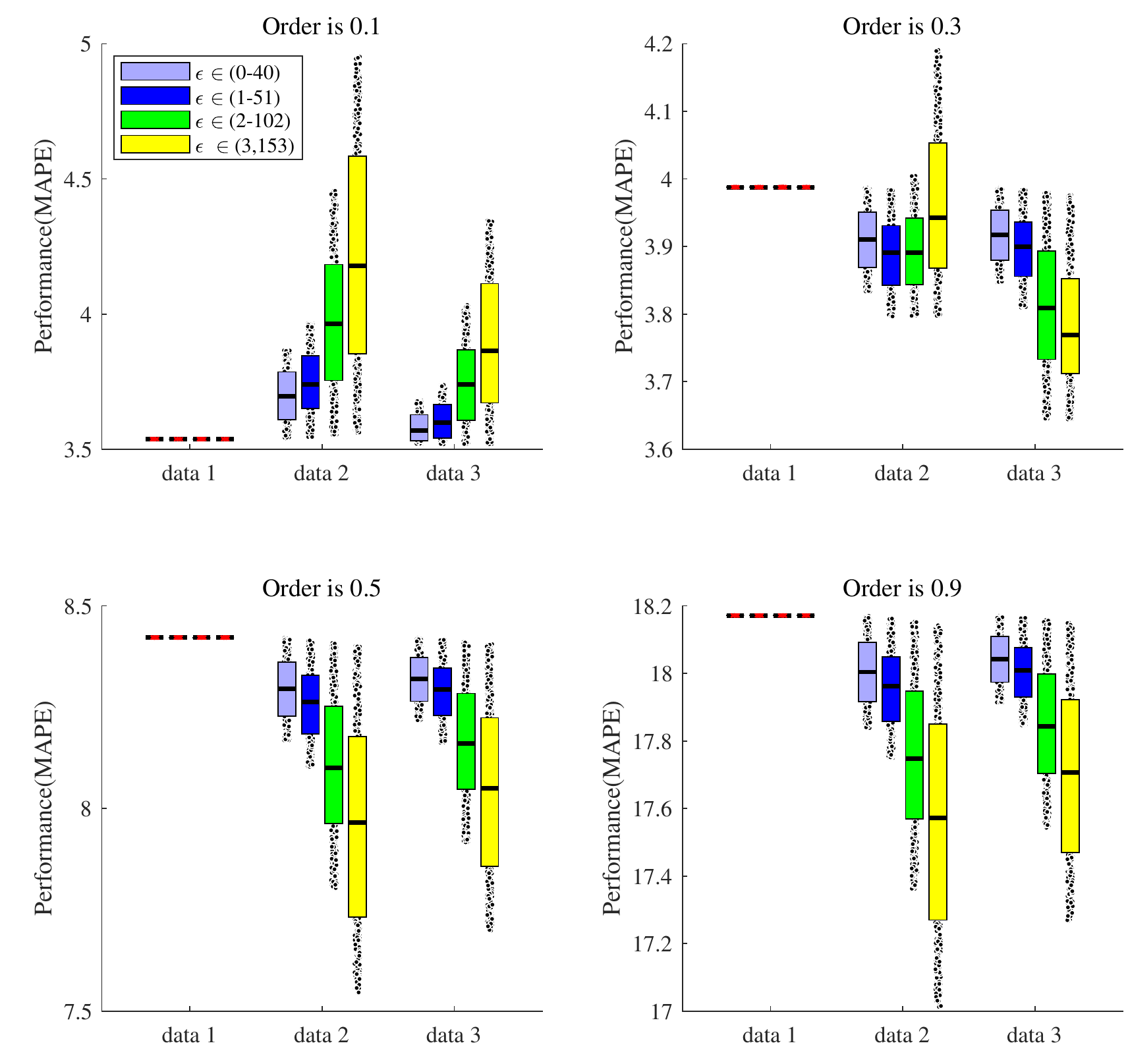}
\caption{Boxplots of the fitting errors with different order and initial value on simulation of GCFM.}
\label{fig:different_order_initial_value}
\end{figure}

\begin{figure}
\centering
\includegraphics[scale=0.4]{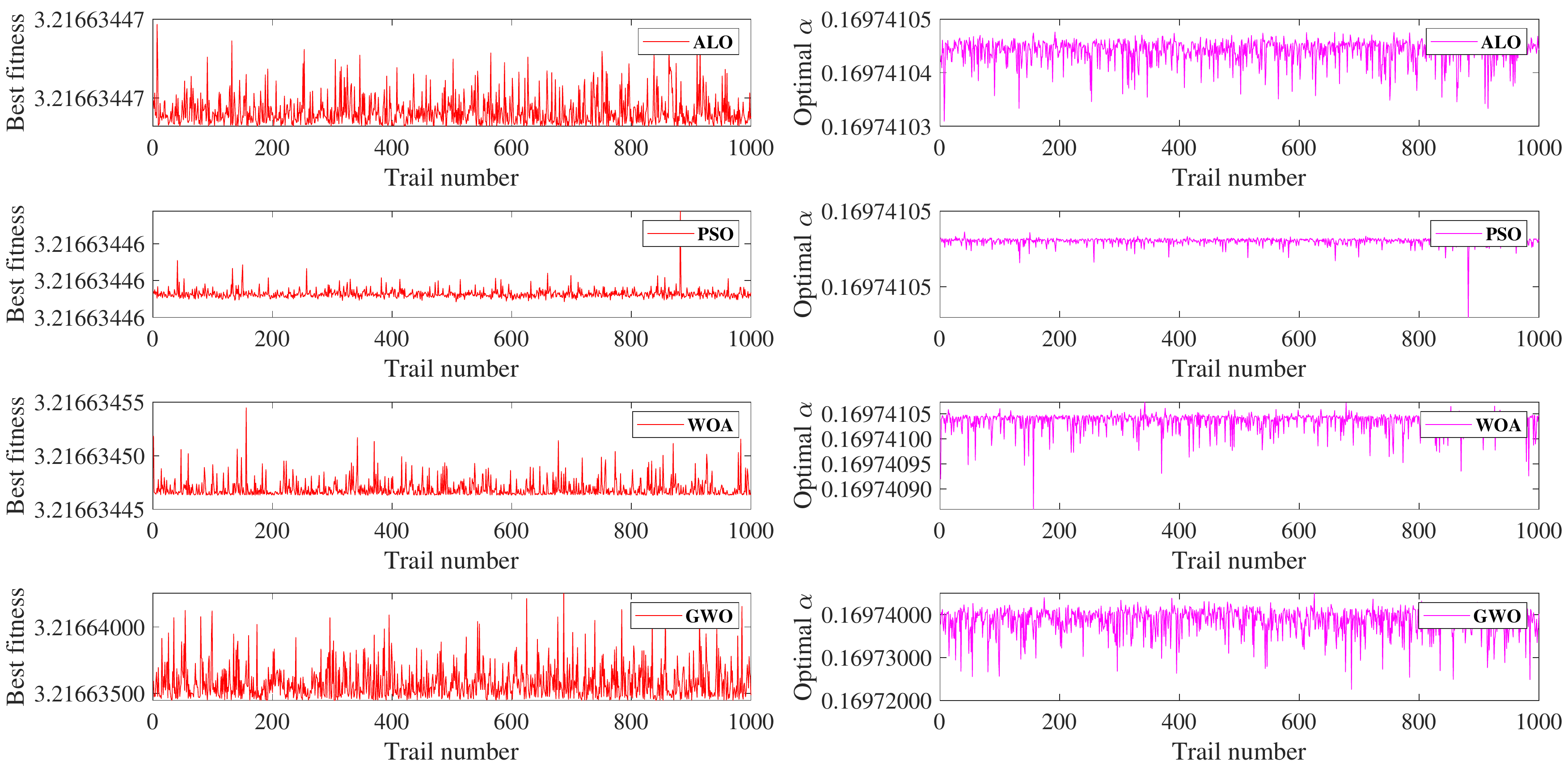}
\caption{Boxplots of the fitting errors with different order and initial value on simulation of GCFM.}
\label{fig:1000times}
\end{figure}

Wu et al. \cite{33} first discusses the information priority in grey model prediction with matrix perturbation bound theory, which is the basis of grey modeling. In this subsection, we will use this technique to analyze the characteristics of GCFGM.
\begin{lemma}[See \cite{63}]
Set $A \in {C^{m \times n}},b \in {C^m},{A^\eta }$ is the generalized inverse matrix of $\text { A, } B=A+W \text { and } c=b+k \in C^{m}$. Then $x$ and $x+h$ satisfy $\|A x-b\|_{2}=\min \text { and }\|B x-c\|_{2}=\min$. If $\operatorname{rank} (A) = \operatorname{rank} (B) = n{\text{ and }}{\left\| {{A^\dagger }} \right\|_2}W{_2} < 1$, then
\begin{eqnarray}
\|h\| \leq \frac{\kappa_{\dagger}}{\gamma_{\dagger}}\left(\frac{\|W\|_{2}}{\|A\|}\|x\|+\frac{\|k\|_{2}}{\|A\|}+\frac{\kappa_{\dagger}}{\gamma_{\dagger}} \frac{\|W\|_{2}}{\|A\|} \frac{\left\|r_{x}\right\|}{\|A\|}\right),
\end{eqnarray}
where $\kappa_{\dagger}=\left\|A^{\dagger}\right\|_{2}\|A\|, \gamma_{\dagger}=1-\left\|A^{\dagger}\right\|_{2}\|W\|_{2}, r_{x}=b-A x$.
\end{lemma}
\begin{theorem} \label{theorem:raodong}
Set $p=1$, we can get the following differential equation $D_{\psi}^{1} x ^{(\alpha)}(t)+ a x ^{(\alpha)}(t)= b$, the corresponding difference equation is ${x^{(\alpha  - 1)}}(k) + \frac{a}{2}\left( {{x^{(\alpha )}}(k) + {x^{(\alpha )}}(k - 1)} \right) = b,k = 1,2,3, \ldots ,n$, where $D_{\psi}^{1} x ^{(\alpha)}(t)$ is equivalent to classic first derivative. Set $x$ is the solution of the GCFGM model,  which satisfies $\min \|B x-Y\|_{2}$. if $\varepsilon$  is a disturbance of original
value $x^{(0)}(k)(k=1,2, \cdots n)$, the perturbation bound of the $x$ is
\begin{eqnarray}
L\left[x^{(0)}(1)\right]=\left|\frac{\varepsilon}{\psi(1, \alpha)}\right| \frac{\kappa_{\dagger}}{\gamma_{\dagger}}\left(\frac{\sqrt{n-1}}{\|B\|}\|x\|+\frac{\kappa_{\dagger}}{\gamma_{\dagger}} \frac{\sqrt{n-1}}{\|B\|} \frac{\left\|r_{x}\right\|}{\|B\|}\right){\text{,   k = 1}}.
\label{x1_set}
\end{eqnarray}
\begin{eqnarray}
L\left[x^{(0)}(k)\right]=\left|\frac{\varepsilon}{\psi(k, \alpha)}\right| \frac{\kappa_{\dagger}}{\gamma_{\dagger}}\left(\frac{\sqrt{n-k+\frac{1}{4}}}{\|B\|}\|x\|+\frac{1}{\|B\|}+\frac{\kappa_{\dagger}}{\gamma_{\dagger}} \frac{\sqrt{n-k+\frac{1}{4}}}{\|B\|} \frac{\left\|r_{x}\right\|}{\|B\|}\right),k = 2,3, \cdots ,n.
\label{pn}
\end{eqnarray}
\textbf{Proof.} if $\varepsilon$ is regarded as a disturbance of $x^{(0)}(1)$, the subsequent is working,
\begin{eqnarray}
Y + \Delta Y = \left( {\begin{array}{*{20}{c}}
  {\frac{{{x^{(0)}}(2)}}{{\psi (2,\alpha)}}} \\
  {\frac{{{x^{(0)}}(3)}}{{\psi (3,\alpha)}}} \\
   \vdots  \\
  {\frac{{{x^{(0)}}(n)}}{{\psi (n,\alpha)}}}
\end{array}} \right) + \left( {\begin{array}{*{20}{c}}
  0 \\
  0 \\
   \vdots  \\
  0
\end{array}} \right),
B + \Delta B = {\text{B + }}\left( {\begin{array}{*{20}{c}}
  { - \frac{\varepsilon }{{\psi (1,\alpha)}}}&0 \\
  { - \frac{\varepsilon }{{\psi (1,\alpha)}}}&0 \\
   \vdots & \vdots  \\
  { - \frac{\varepsilon }{{\psi (1,\alpha)}}}&0
\end{array}} \right).
\end{eqnarray}
Therefore, $\|\Delta Y\|_{2}=0, \|\Delta B\|_{2}=\sqrt{n-1}\left|\frac{\varepsilon}{\psi(1, r)}\right|$,so the the perturbation bound can be  defined as
\begin{eqnarray}
\|\Delta x\| \leq \frac{\kappa_{\dagger}}{\gamma_{\dagger}}\left(\frac{\|\Delta B\|_{2}}{\|B\|}\|x\|+\frac{\|\Delta Y\|_{2}}{\|B\|}+\frac{\kappa_{\dagger}}{\gamma_{\dagger}} \frac{\|\Delta B\|_{2}}{\|B\|} \frac{\left\|r_{x}\right\|}{\|B\|}\right).
\end{eqnarray}
So, Eq. (\ref{x1_set}) is proved. If $\varepsilon$ is regarded as a disturbance of $x^{(0)}(2)$, then
\begin{eqnarray}
Y + \Delta Y = \left( {\begin{array}{*{20}{c}}
  {\frac{{{x^{(0)}}(2)}}{{\psi (2,\alpha)}}} \\
  {\frac{{{x^{(0)}}(3)}}{{\psi (3,\alpha)}}} \\
   \vdots  \\
  {\frac{{{x^{(0)}}(n)}}{{\psi (n,\alpha)}}}
\end{array}} \right) + \left( {\begin{array}{*{20}{c}}
  {\frac{\varepsilon }{{\psi (2,\alpha)}}} \\
  0 \\
   \vdots  \\
  0
\end{array}} \right)B + \Delta B = B{\text{ + }}\left( {\begin{array}{*{20}{c}}
  { - \frac{\varepsilon }{{2\psi (2,\alpha)}}}&0 \\
  { - \frac{\varepsilon }{{\psi (2,\alpha)}}}&0 \\
   \vdots & \vdots  \\
  { - \frac{\varepsilon }{{\psi (2,\alpha)}}}&0
\end{array}} \right).
\end{eqnarray}
The perturbation bound can be expressed as
\begin{eqnarray}
L\left[x^{(0)}(2)\right]=\left|\frac{\varepsilon}{\psi(2, r)}\right| \frac{\kappa_{\dagger}}{\gamma_{\dagger}}\left(\frac{\sqrt{n-\frac{7}{4}}}{\|B\|}\|x\|+\frac{1}{\|B\|}+\frac{\kappa_{\dagger}}{\gamma_{\dagger}} \frac{\sqrt{n-\frac{7}{4}}}{\|B\|} \frac{\left\|r_{x}\right\|}{\|B\|}\right).
\end{eqnarray}
 If $\varepsilon$ is regarded as a disturbance of $x^{(0)}(n)$, we can get
\begin{eqnarray}
L\left[x^{(0)}(n)\right]=\left|\frac{\varepsilon}{\psi(n, r)}\right| \frac{\kappa_{\dagger}}{\gamma_{\dagger}}\left(\frac{1}{2\|B\|}\|x\|+\frac{1}{\|B\|}+\frac{\kappa_{\dagger}}{\gamma_{\dagger}} \frac{1}{2\|B\|} \frac{\left\|r_{x}\right\|}{\|B\|}\right).
\end{eqnarray}
Without loss of generality, when $k=2,3, \cdots, n$, we can easily obtain Eq. (\ref{pn}) using the same method as above.
\end {theorem}
In the above analysis, we use perturbation boundary theory to prove the stability of the GCFGM(1,1) model with  different disturbed raw data. Through  \textcolor{blue}{Theorem} \ref{theorem:raodong}, the following conclusion is working, the sample size $n$ is an increase function of $L\left[ {{x^{(0)}}(k)} \right]k = 1,2,3,...,n$. Therefore, in order to increase the stability of GCFGM model, we should use less data in actual modeling background. In the following study, we will explore the influence of initial value on GCFGM(1,1).
\begin{theorem} \label{theorem:initial}
Set $X^{(0)}=\left\{x^{(0)}(1), x^{(0)}(2), \cdots, x^{(0)}(n)\right\}$ is the raw nonnegative data. if $\varepsilon$ is regarded as a disturbance of $x^{(0)}(1)$ and $\psi ({\text{1}},\alpha ){\text{ = 1}}$. $x^{(0)}(1)+\varepsilon$  dose not cause the changing of GCFGM's simulative value $\hat{X}^{(0)}=\left\{\hat{x}^{(0)}(2), \cdots, \hat{x}^{(0)}(n), \hat{x}^{(0)}(n+1), \cdots\right\}$.
\end{theorem}
\textbf{Proof.} When ${x^{(0)}}(1) + \varepsilon$  exists, we have $X^{(\alpha)}+\varepsilon=\left\{x^{(0)}(1)+\varepsilon, x^{(\alpha)}(2)+\varepsilon, \cdots, x^{(\alpha)}(n)+\varepsilon\right\}$. The complete proof process is similar to \textcolor{blue}{Theorem 1} of Reference \cite{64}. In order to verify the validity of this conclusion, we furniture our results by illustrative numerical examples for the GCFGM (Sets 1 is the  electricity consumption of Jiangsu province in China published by China¡¯s National Statistics Bureau (http://www.stats.gov.cn/english/)) and choose \textcolor {blue}{Eq. ({\ref{Recomand_GCFA}})} as the accumulation of GCFM.
\begin{table}[htbp]\scriptsize
  \centering
  \caption{The simulation results of GCFGM with different initial values}

    \begin{tabular}{lrrrrrrrrr}
    \toprule
     Sets 1 & \multicolumn{1}{l}{Sets 2} & \multicolumn{1}{l}{Model 1} & \multicolumn{1}{l}{APE(\%)} & \multicolumn{1}{l}{Model 2} & \multicolumn{1}{l}{APE(\%)} & \multicolumn{1}{l}{Model 3} & \multicolumn{1}{l}{APE(\%)} & \multicolumn{1}{l}{Model 4} & \multicolumn{1}{l}{APE(\%)} \\
    \midrule
    \multicolumn{1}{r}{971.34} & 1071.34  & 971.34  & 0.00  & 1071.34  & 0.00  & 971.34 & 0.00  & 1071.34 & 0.00  \\
    \multicolumn{1}{r}{1078.44} & 1078.44 & 1431.75  & 32.76  & 1431.75  & 32.76  & 918.4131 & 14.84  & 918.4131 & 14.84  \\
    \multicolumn{1}{r}{1245.14} & 1245.14 & 1542.83  & 23.91  & 1542.83  & 23.91  & 1253.323 & 0.66  & 1253.323 & 0.66  \\
    \multicolumn{1}{r}{1505.13} & 1505.13 & 1721.41  & 14.37  & 1721.41  & 14.37  & 1609.24 & 6.92  & 1609.24 & 6.92  \\
    \multicolumn{1}{r}{1820.08} & 1820.08 & 1937.80  & 6.47  & 1937.80  & 6.47  & 1955.808 & 7.46  & 1955.808 & 7.46  \\
    \multicolumn{1}{r}{2193.45} & 2193.45 & 2176.49  & 0.77  & 2176.49  & 0.77  & 2291.823 & 4.48  & 2291.823 & 4.48  \\
    \multicolumn{1}{r}{2569.75} & 2569.75 & 2429.99  & 5.44  & 2429.99  & 5.44  & 2618.384 & 1.89  & 2618.384 & 1.89  \\
    \multicolumn{1}{r}{2952.02} & 2952.02 & 2695.04  & 8.71  & 2695.04  & 8.71  & 2936.48 & 0.53  & 2936.48 & 0.53  \\
    \multicolumn{1}{r}{3118.32} & 3118.32 & 2970.53  & 4.74  & 2970.53  & 4.74  & 3246.873 & 4.12  & 3246.873 & 4.12  \\
    \multicolumn{1}{r}{3313.99} & 3313.99 & 3256.41  & 1.74  & 3256.41  & 1.74  & 3550.162 & 7.13  & 3550.162 & 7.13  \\
    \multicolumn{1}{r}{3864.37} & 3864.37 & 3553.20  & 8.05  & 3553.20  & 8.05  & 3846.826 & 0.45  & 3846.826 & 0.45  \\
    \multicolumn{1}{r}{4281.62} & 4281.62 & 3861.63  & 9.81  & 3861.63  & 9.81  & 4137.264 & 3.37  & 4137.264 & 3.37  \\
    \multicolumn{1}{r}{4580.90} & 4580.90 & 4182.58  & 8.70  & 4182.58  & 8.70  & 4421.811 & 3.47  & 4421.811 & 3.47  \\
    \multicolumn{1}{r}{4956.60} & 4956.60 & 4516.95  & 8.87  & 4516.95  & 8.87  & 4700.753 & 5.16  & 4700.753 & 5.16  \\
    \multicolumn{1}{r}{5012.54} & 5012.54 & 4865.69  & 2.93  & 4865.69  & 2.93  & 4974.338 & 0.76  & 4974.338 & 0.76  \\
    \multicolumn{1}{r}{5114.70} & 5114.70 & 5229.73  & 2.25  & 5229.73  & 2.25  & 5242.787 & 2.50  & 5242.787 & 2.50  \\
    \multicolumn{1}{r}{5458.95} & 5458.95 & 5610.04  & 2.77  & 5610.04  & 2.77  & 5506.292 & 0.87  & 5506.292 & 0.87  \\
    \multicolumn{1}{r}{5807.89} & 5807.89 & 6007.60  & 3.44  & 6007.60  & 3.44  & 5765.029 & 0.74  & 5765.029 & 0.74  \\
    \multicolumn{1}{r}{6128.27} & 6128.27 & 6423.40  & 4.82  & 6423.40  & 4.82  & 6019.155 & 1.78  & 6019.155 & 1.78  \\
    \multicolumn{1}{r}{6264.36} & 6264.36 & 6858.47  & 9.48  & 6858.47  & 9.48  & 6268.813 & 0.07  & 6268.813 & 0.07  \\
    MAPE(\%) &       &       & 8.42  &       & 8.42  &       & 3.54  &       & 3.54  \\
    \bottomrule
    \end{tabular}%
  \label{tab:initial_table}%
\end{table}%
From \textcolor{blue}{Table} \ref{tab:initial_table}, we can easily see that although the initial value has changed under the action of different orders, the simulated value of GCFGM has not changed. The initial value of Sets 2 is different from of Sets 1, and other data are consistent. Model 1 and Model 3 are different models built with Sets 1, their cumulative orders are 0.5 and 0.1 respectively. Model 2 and Model 4 are different models built with Sets 2, and their cumulative orders are 0.5 and 0.1 respectively. In order to further describe the influence of different orders and different initial values on fitting results of GCFM, four cumulative orders and different initial values can be employed to observe the fitting results of the raw samples. We change first three values of the model (${x^{(0)}}(1),{x^{(0)}}(2),{x^{(0)}}(3)$) into disturbance values  ($x^{(0)}(1)+\epsilon, x^{(0)}(2)+\epsilon, x^{(0)}(3)+\epsilon$) in four different intervals , and generate 500 disturbance values in each interval with different $\epsilon$ . By substituting different disturbance values into GCFM, different fitting errors are obtained. It can be seen from \textcolor{blue}{Figure} \ref{fig:different_order_initial_value}, Under the influence of different cumulative order and disturbance intensity, the change of initial value do not affect the fitting values of the model, while the change of second value and the third value will directly affect the fitting results of the model.
In this numerical example, the another aim is to compare
the performances of the (WOA\cite{57}, ALO\cite{58}, GWO\cite{59}, PSO). we use the four algorithms to search the minimum MAPE and the corresponding order $\alpha$ of the GCFGM(1,1) model among the 1000 trails. The optimal orders of th e model and MAPE in each trail are presented in textcolor{blue}{Fig.} \ref{fig:1000times}. It can be seen clearly in textcolor{blue}{Fig.} \ref{fig:1000times}that PSO is more stable than the other three optimizer. According to the above analysis, the PSO should be employed to construct GCFGM(1,1) model in the engineering  application when we need more stable output of the model.
\section{Application}
\label{s4}
In this section, we use the GCFGM(1,1) model to evaluate China's total energy consumption (10,000 tons of standard coal) and natural gas consumption (10,000 tons of standard coal) (data sets were published by China's National Statistics Bureau (http://www.stats.gov.cn/english/)). For simple consideration, we choose the order of the differential equation as 1. In order to comprehensively utilize the characteristics of different accumulations, we choose \textcolor{blue}{Eq.} (\ref{suggestion_accumulation}) as the fractional accumulation. To verify the effectiveness of the proposed model, we use several other representative models (CFGM(1,1)\cite{34}, GM(1,1)\cite{14}, DGM(1,1)\cite{61}) to compare with our GCFGM(1,1). We use MAPE as the evaluation standard of the model, and their definitions is
\begin{eqnarray}
\operatorname{APE} (k) = \left| {\frac{{{x^{(0)}}(k) - {{\hat x}^{(0)}}(k)}}{{{x^{(0)}}(k)}}} \right| \times 100\% ,k = 2,3, \ldots ,n
\end{eqnarray}
\begin{eqnarray}
\text{MAPE}=\frac{1}{n}\sum_{i=1}^n\frac{\left|\widehat{x}^{(0)}(k)-x^{(0)}(k)\right|}{x^{(0)}(k)}\times100\%
\end{eqnarray} and choose \textcolor {blue}{Eq. ({\ref{Recomand_GCFA}})} as the accumulation of GCFM. Compared with PSO, although the other three algorithms (PSO, ALO, GWO,) have been proved to have excellent characteristics, they have also been widely applied to complex problems in various fields. But our problem is relatively simple, we only need to search for one parameter. Through the above analysis, we found that PSO has better stability. Here we consider more about the stability of GCFGM(1,1), so in the stage of application, we consider using PSO algorithm.

\begin{figure}
\centering
\includegraphics[scale=0.6]{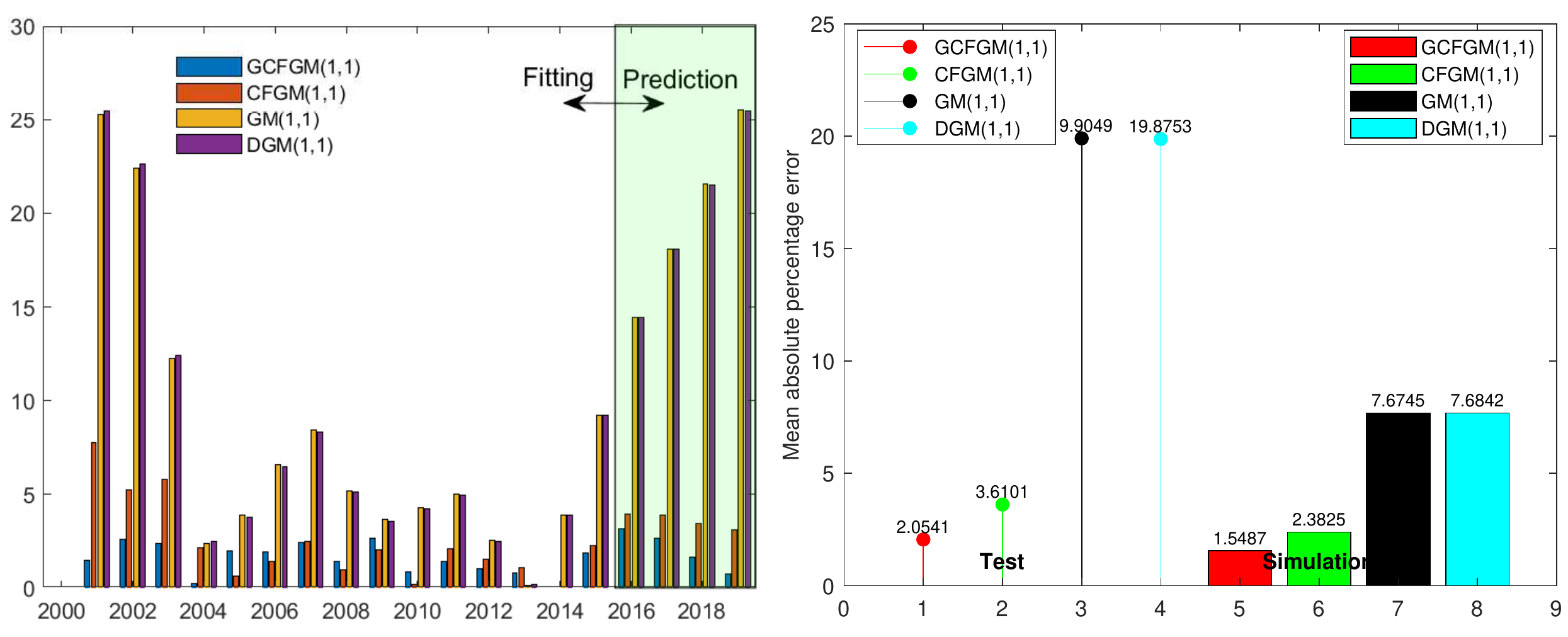}
\caption{APEs of four prediction models, GCFGM(1,1), DGM(1,1), GM(1,1), CFGM(1,1) in Case 1 (Left). MAPEs of four prediction models (Right) in Case 1.}
\label{case1:zhe}
\end{figure}

\textbf{Case 1.} In this case, we use four models, (GCFGM(1,1), CFGM, GM(1,1), DGM(1,1)), to predict China's overall energy consumption. The data from 2010 to 2015 are used as the training samples to build the models, and the data from 2016 to 2019 are used as the test samples. Finally, we calculate MAPEs of the models on the training set and the test set. The results can be seen in \textcolor{blue} {Table} \ref{case1:simu} and \textcolor{blue} {Figure} \ref{case1:zhe}.
\begin{table}[h]\scriptsize
  \centering
  \caption{Numerical results by GCFGM(1,1), CFGM(1,1), GM(1,1), DGM(1,1) in Case 1.}
    \begin{tabular}{lrrrrrrrrr}
    \toprule
    Year  & \multicolumn{1}{l}{True value} & \multicolumn{1}{l}{GCFGM(1,1)} & \multicolumn{1}{l}{Error(\%)} & \multicolumn{1}{l}{CFGM(1,1)} & \multicolumn{1}{l}{Error(\%)} & \multicolumn{1}{l}{GM(1,1)} & \multicolumn{1}{l}{Error(\%)} & \multicolumn{1}{l}{DGM(1,1)} & \multicolumn{1}{l}{Error(\%)} \\
    \midrule
    \multicolumn{1}{r}{2000} & 146964.00 & 146964 & 0     & 146964 & 0     & 146964 & 0     & 146964 & 0 \\
    \multicolumn{1}{r}{2001} & 155547.00 & 153215 & 1.499193 & 143444.7 & 7.780508 & 194808.3 & 25.24076 & 195148.2 & 25.45932 \\
    \multicolumn{1}{r}{2002} & 169577.00 & 174011.2 & 2.614887 & 178497.3 & 5.260306 & 207586.3 & 22.41416 & 207923.5 & 22.61303 \\
    \multicolumn{1}{r}{2003} & 197083.00 & 201730.1 & 2.357929 & 208504.8 & 5.79542 & 221202.4 & 12.23821 & 221535.1 & 12.40701 \\
    \multicolumn{1}{r}{2004} & 230281.00 & 229724.2 & 0.241785 & 235262.4 & 2.163162 & 235711.7 & 2.358297 & 236037.8 & 2.499905 \\
    \multicolumn{1}{r}{2005} & 261369.00 & 256213 & 1.972672 & 259699.2 & 0.638866 & 251172.7 & 3.901114 & 251489.9 & 3.779747 \\
    \multicolumn{1}{r}{2006} & 286467.00 & 280880.3 & 1.950208 & 282371.2 & 1.429771 & 267647.8 & 6.569408 & 267953.6 & 6.462668 \\
    \multicolumn{1}{r}{2007} & 311442.00 & 303812.4 & 2.449779 & 303642.4 & 2.504345 & 285203.6 & 8.424818 & 285495 & 8.33123 \\
    \multicolumn{1}{r}{2008} & 320611.00 & 325179.9 & 1.425075 & 323767 & 0.984368 & 303910.9 & 5.208843 & 304184.9 & 5.123388 \\
    \multicolumn{1}{r}{2009} & 336126.00 & 345150 & 2.684719 & 342930.6 & 2.024413 & 323845.2 & 3.653619 & 324098.2 & 3.578366 \\
    \multicolumn{1}{r}{2010} & 360648.00 & 363866.5 & 0.892424 & 361273.8 & 0.173524 & 345087.1 & 4.314693 & 345315.1 & 4.25148 \\
    \multicolumn{1}{r}{2011} & 387043.00 & 381449.8 & 1.445117 & 378906.2 & 2.102287 & 367722.4 & 4.991855 & 367921 & 4.940531 \\
    \multicolumn{1}{r}{2012} & 402138.00 & 398000.6 & 1.028843 & 395915.2 & 1.547422 & 391842.3 & 2.560238 & 392006.8 & 2.519333 \\
    \multicolumn{1}{r}{2013} & 416913.00 & 413604.2 & 0.793635 & 412371.7 & 1.089261 & 417544.3 & 0.151431 & 417669.4 & 0.181418 \\
    \multicolumn{1}{r}{2014} & 428333.99 & 428333.4 & 0.000135 & 428334.4 & 8.42E-05 & 444932.2 & 3.87507 & 445011.9 & 3.893668 \\
    \multicolumn{1}{r}{2015} & 434112.78 & 442251.2 & 1.874714 & 443852.2 & 2.243512 & 474116.6 & 9.21507 & 474144.4 & 9.221478 \\
    MAPE  &       &       & 1.548741 &       & 2.382483 &       & 7.674506 &       & 7.684171 \\
    \multicolumn{1}{r}{2016} & 441491.81 & 455412.5 & 3.153109 & 458966.7 & 3.958146 & 505215.2 & 14.43365 & 505184 & 14.4266 \\
    \multicolumn{1}{r}{2017} & 455826.92 & 467866.1 & 2.64118 & 473713.6 & 3.924001 & 538353.7 & 18.10484 & 538255.7 & 18.08335 \\
    \multicolumn{1}{r}{2018} & 471925.15 & 479655.2 & 1.637983 & 488123.5 & 3.4324 & 573665.8 & 21.55864 & 573492.4 & 21.52189 \\
    \multicolumn{1}{r}{2019} & 487000.00  & 490818.5 & 0.784089 & 502223.3 & 3.12593 & 611294.1 & 25.52241 & 611035.8 & 25.46936 \\
    MAPE &       &       & 2.05409 &       & 3.610119 &       & 19.90489 &       & 19.8753 \\
    \bottomrule
    \end{tabular}%
 \label{case1:simu}
\end{table}%
% Table generated by Excel2LaTeX from sheet 'ÕûÀícase1'
% Table generated by Excel2LaTeX from sheet 'ÕûÀícase1'
\begin{table}[htbp]\scriptsize
  \centering
  \caption{The fitness (MAPE) and order searched by the PSO algorithms in Case 1.}
\setlength{\tabcolsep}{16mm}{%7¿ÉËæ»úÉèÖ㬵÷Õûµ½ÊʺÏ×Ô¼ºµÄ´óСΪֹ
    \begin{tabular}{lrr}
    \toprule
          & \multicolumn{1}{l}{PSO-GCFGM} & \multicolumn{1}{l}{PSO-CFGM} \\
    \midrule
    Order & 0.3228 & 0.4631 \\
    MAPE  & 1.5487 & 2.3825 \\
    \bottomrule
    \end{tabular}}%
  \label{tab:PSO_case1}%
\end{table}%
It can be seen from \textcolor{blue}{Table} \ref{case1:simu} that in fitting stage, MAPEs of GCFGM(1,1),  CFGM(1,1), GM(1,1) and DGM(1,1) are 1.548741\%, 2.382483\%, 7.674506\% and 7.684171\%, respectively. In prediction stage, MAPEs are 1.548741\%, 2.382483\%, 7.674506\%, 7.684171\%, respectively. It can be found that the error of GCFGM(1,1) is the smallest one in both fitting stage and prediction stage. This verifies that GCFGM(1,1) has certain advantages. The fitness (MAPE) and order searched by the PSO algorithm are shown in \textcolor{blue}{Table} \ref{tab:PSO_case1}, we can see that after the optimization of PSO, the order of GCFGM (1,1) model is 0.3228, and the corresponding MAPE is 1.5487. The order of CFGM(1,1) model is 0.4631, and the corresponding MAPE is 2.3825.

\textbf{Case 2.} Forecasting China's natural gas consumption. The prediction of natural gas is of great significance and can provide important suggestions to decision makers. In this Case, we use the model to fit China's natural gas consumption data from 2000 to 2015 and the data from 2016 to 2019 to test the established model, and calculate MAPE of the fitting and predicting stages respectively. It can be found from \textcolor{blue}{Tabel} \ref{case2:simu} that in fitting stage,  MAPEs of GCFGM(1,1), CFGM(1,1), GM(1,1), DGM(1,1) are 4.125082\%, 5.621349\%, 9.389527\%, 9.699885\% respectively. The MAPE in prediction phase is 8.7433\%, 9.319867\%, 21.20638\%, 21.80221\%. It can be seen from \textcolor{blue}{Tabel} \ref{case2:simu} that in this case, compared with the other three models, the output of GCFGM is closest to the real value regardless of the fitting order and the prediction stage. Like Case 1,  The fitness (MAPE) and order searched by the PSO algorithm are shown in \textcolor{blue}{Table} \ref{Case2_order}, we can see that after the optimization of PSO, the order of GCFGM(1,1) model is 0.4382, and the corresponding MAPE is 4.1251. The order of CFGM(1,1) model is 0.58628, and the corresponding MAPE is 5.6213.

From these two cases, we can see that it is possible to improve the fitting and prediction accuracy of the model by reconstructing the grey prediction model with GCFA and GCFD. In practical modeling problems, we can flexibly adjust our accumulation types of accumulation according to the establishment principles of GCFA and GCFD when the higher accuracy is needed in fitting and
forecasting.
\begin{table}\scriptsize
  \centering
  \caption{Numerical results by GCFGM(1,1), CFGM(1,1), GM(1,1), DGM(1,1) in Case 2.}
    \begin{tabular}{lrrrrrrrrr}
    \toprule
    Year  & \multicolumn{1}{l}{True value} & \multicolumn{1}{l}{GCFGM(1,1)} & \multicolumn{1}{l}{Error(\%)} & \multicolumn{1}{l}{CFGM} & \multicolumn{1}{l}{Error(\%)} & \multicolumn{1}{l}{GM(1,1)} & \multicolumn{1}{l}{Error(\%)} & \multicolumn{1}{l}{DGM(1,1)} & \multicolumn{1}{l}{Error(\%)} \\
    \midrule
    \multicolumn{1}{r}{2000} & 3233.21 & 3233.21 & 0     & 3233.21 & 0     & 3233.21 & 0     & 3233.21 & 0 \\
    \multicolumn{1}{r}{2001} & 3733.13 & 3356.357 & 10.09268 & 2998.57 & 19.67679 & 4217.306 & 12.9697 & 4237.064 & 13.49897 \\
    \multicolumn{1}{r}{2002} & 3900.27 & 3900.32 & 0.001285 & 3900.292 & 0.000564 & 4834.06 & 23.94168 & 4856.776 & 24.52409 \\
    \multicolumn{1}{r}{2003} & 4532.91 & 4691.734 & 3.503796 & 4831.9 & 6.59599 & 5541.011 & 22.2396 & 5567.126 & 22.81572 \\
    \multicolumn{1}{r}{2004} & 5296.46 & 5651.381 & 6.701094 & 5828.317 & 10.04174 & 6351.349 & 19.91688 & 6381.372 & 20.48372 \\
    \multicolumn{1}{r}{2005} & 6272.86 & 6747.777 & 7.570974 & 6912.479 & 10.1966 & 7280.194 & 16.05861 & 7314.71 & 16.60884 \\
    \multicolumn{1}{r}{2006} & 7734.61 & 7973.455 & 3.087998 & 8103.32 & 4.767016 & 8344.877 & 7.890084 & 8384.557 & 8.403095 \\
    \multicolumn{1}{r}{2007} & 9343.26 & 9332.969 & 0.110147 & 9418.625 & 0.806625 & 9565.263 & 2.376079 & 9610.879 & 2.864302 \\
    \multicolumn{1}{r}{2008} & 10900.77 & 10837.15 & 0.583623 & 10876.33 & 0.224174 & 10964.12 & 0.58118 & 11016.56 & 1.062248 \\
    \multicolumn{1}{r}{2009} & 11764.41 & 12500.46 & 6.256613 & 12495.27 & 6.212442 & 12567.56 & 6.826925 & 12627.84 & 7.339358 \\
    \multicolumn{1}{r}{2010} & 14425.92 & 14339.85 & 0.596613 & 14295.61 & 0.903272 & 14405.48 & 0.141662 & 14474.79 & 0.33874 \\
    \multicolumn{1}{r}{2011} & 17803.98 & 16374.31 & 8.030042 & 16299.3 & 8.451343 & 16512.2 & 7.255592 & 16591.86 & 6.808119 \\
    \multicolumn{1}{r}{2012} & 19302.62 & 18624.79 & 3.511579 & 18530.33 & 4.000938 & 18927 & 1.945948 & 19018.58 & 1.471486 \\
    \multicolumn{1}{r}{2013} & 22096.39 & 21114.28 & 4.444682 & 21015.08 & 4.893586 & 21694.96 & 1.816739 & 21800.24 & 1.340279 \\
    \multicolumn{1}{r}{2014} & 23986.7 & 23867.93 & 0.49514 & 23782.64 & 0.850741 & 24867.71 & 3.672901 & 24988.73 & 4.177449 \\
    \multicolumn{1}{r}{2015} & 25178.54 & 26913.33 & 6.889961 & 26865.1 & 6.698415 & 28504.45 & 13.20932 & 28643.58 & 13.76186 \\
    Mape  &       &       & 4.125082 &       & 5.621349 &       & 9.389527 &       & 9.699885 \\
    \multicolumn{1}{r}{2016} & 26931 & 30280.68 & 12.43801 & 30297.99 & 12.50229 & 32673.05 & 21.32135 & 32832.98 & 21.91518 \\
    \multicolumn{1}{r}{2017} & 31452.06 & 34003.08 & 8.110829 & 34120.58 & 8.484397 & 37451.28 & 19.07418 & 37635.12 & 19.65867 \\
    \multicolumn{1}{r}{2018} & 35866.31 & 38116.84 & 6.27478 & 38376.34 & 6.998287 & 42928.3 & 19.68975 & 43139.62 & 20.27894 \\
    \multicolumn{1}{r}{2019} & 39447.00 & 42661.77 & 8.149585 & 43113.4 & 9.294497 & 49206.29 & 24.74026 & 49449.2 & 25.35606 \\
    MAPE  &       &       & 8.7433 &       & 9.319867 &       & 21.20638 &       & 21.80221 \\
    \bottomrule
    \end{tabular}%
    \label{case2:simu}
\end{table}%
% Table generated by Excel2LaTeX from sheet ';˒case2'
\begin{table}[htbp]\scriptsize
  \centering
  \caption{The fitness (MAPE) and order searched by the PSO algorithms in Case 2.}
  \setlength{\tabcolsep}{16mm}{%7 ¿ÉËæ»úÉèÖ㬵÷Õûµ½ÊʺÏ×Ô¼ºµÄ´óСΪֹ
    \begin{tabular}{lrr}
    \toprule
          & \multicolumn{1}{l}{PSO-GCFGM} & \multicolumn{1}{l}{PSO-CFGM} \\
    \midrule
    Order & 0.4382 & 0.58628 \\
    MAPE  & 4.1251 & 5.6213 \\
    \bottomrule
    \end{tabular}}%
  \label{Case2_order}%
\end{table}%
\begin{figure}
\centering
\includegraphics[scale=0.6]{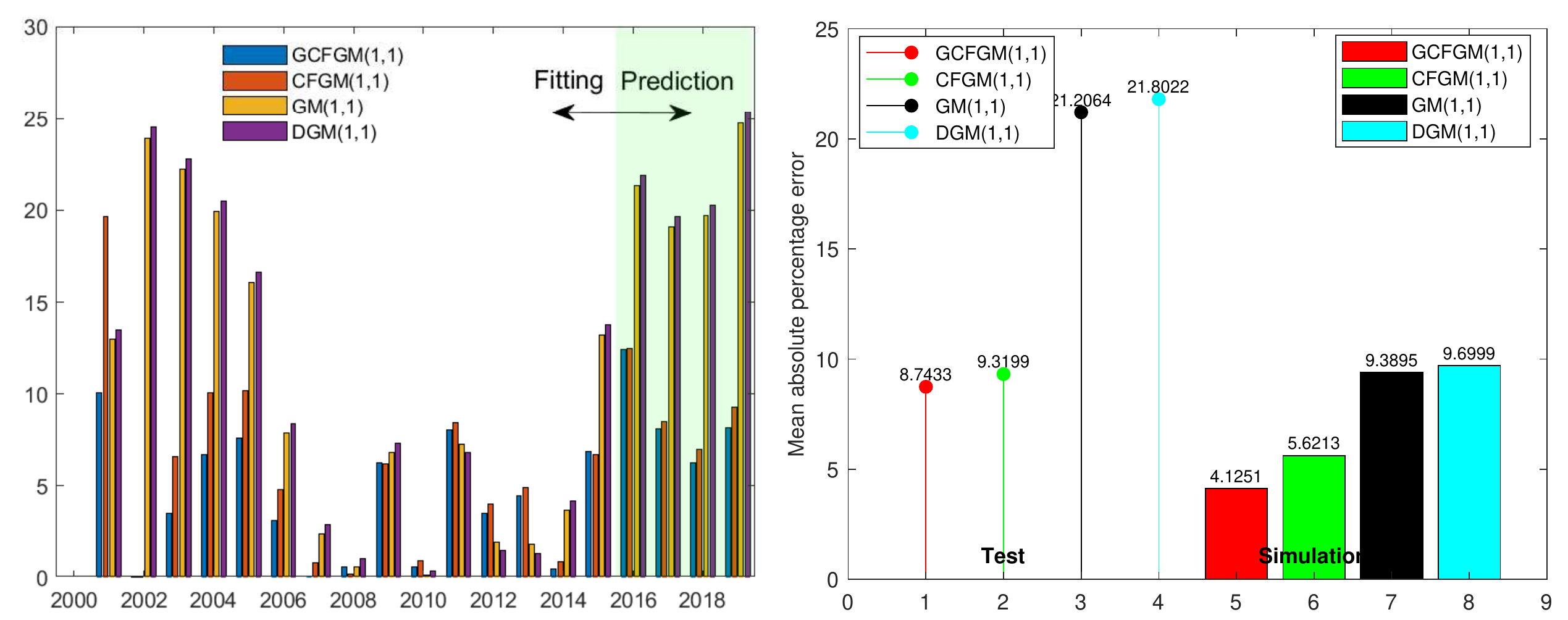}
\caption{APEs of four prediction models, GCFGM(1,1), DGM(1,1), GM(1,1), CFGM(1,1) in Case 2 (Left). MAPEs of four prediction models (Right) in Case 2.}
\label{case2:zhe}
\end{figure}
\section{Conclusion}
\label{s5}
Grey system theory is an important modeling tool, which has successfully solved many engineering and social problems. But we hope to understand deeper meaning of the theory. In this paper, we explained   the important role of cumulative generation in grey system models from the perspective of complex networks. We also explained the physics meaning of the grey model with conformable derivative, and proposed a new grey model. The main contribution of our work are  as follows:

(1) For the first time, we explained an important discovery based on the perspective of complex networks that the effect of cumulative generation can enhance the efficiency of information transmission.

(2) We propose a generalized conformable accumulation and difference is proposed and explain the physical meaning of them.

(3) We propose a new grey prediction model, GCFGM(1,1) based on GCFA and GCFD, and use four optimizers to search the order $\alpha$ of the model. By two practical examples, we verify the effectiveness of our model.

Experiments shows that GCFGM(1,1) has some good characteristics and better modeling accuracy compared to traditional models. At the same time, in this article, we give the important role of accumulation in grey system theory. In the future, a grey prediction model with ability to capture non-linear characteristics of raw data can be constructed. We also need to find a way to select an appropriate function to improve our modeling accuracy.

\noindent\textbf{Acknowledgements}
The work in this paper was supported by grants from the National Natural Science Foundation of China [Grant No.41631175, 61702068, 62007028], the Key Project of Ministry of Education for the 13th 5-years Plan of National Education Science of China [Grant No.DCA170302], the Social Science Foundation of Jiangsu Province of China [Grant No.15TQB005], the Priority Academic Program Development of Jiangsu Higher Education Institutions [Grant No.1643320H111] and the Fundamental Research Funds for the Central Universities of China (Grant No. 2019YBZZ062).

%\noindent\textbf{References}
%\begin{thebibliography}{}
%\bibitem{R1}
%BP, 2017. Energy Outlook-2017 edition. BP, UK.
%\bibitem{R2}
%F. Liang, M. Ryvak, S. Sayeed,  N. Zhao, The role of natural gas as a primary fuel in the near future, including comparisons of acquisition, transmission and waste handling costs of as with competitive alternatives. Chemistry Central Journal 6(S1)(2012) S4.
%\end{thebibliography}
\end{document}